\documentclass[twocolumn,showpacs,preprintnumbers,amsmath,amssymb]{revtex4}

\usepackage{graphicx}% Include figure files
\usepackage{dcolumn}% Align table columns on decimal point
\usepackage{bm}% bold math
\usepackage{mathrsfs}
\usepackage{color}

\begin{document}

\title{Cluster-shell competition and its effect on the $E0$ transition probability in $^{20}$Ne}% 

\author{N. Itagaki$^1$ and  H. Matsuno$^2$}

\affiliation{
$^1$Yukawa Institute for Theoretical Physics, Kyoto University,
Kitashirakawa Oiwake-Cho, Kyoto 606-8502, Japan\\
$^2$Department of Physics, Kyoto University,
Kitashirakawa Oiwake-Cho, Kyoto 606-8502, Japan\\
}

\date{\today}

\begin{abstract}
$^{20}$Ne has been known as a typical example of a nucleus with $\alpha$ cluster structure
($^{16}$O+$\alpha$ structure).
However according to the spherical shell model, the spin-orbit interaction acts attractively
for four nucleons outside of the $^{16}$O core,
and this spin-orbit effect cannot be taken into account
in the simple $\alpha$ cluster models.
We investigate how the $\alpha$ cluster structure 
competes with independent particle motions of these four nucleons.
The antisymmetrized quasi-cluster model (AQCM) is a method 
to describe a transition from the $\alpha$ cluster wave function to the $jj$-coupling shell model
wave function. In this model, the cluster-shell transition is characterized by only two parameters; 
$R$ representing the distance between clusters 
and $\Lambda$ describing the breaking of $\alpha$ clusters,
and the contribution of the spin-orbit interaction, very important in the 
$jj$-coupling shell model,
can be taken into account by changing $\alpha$ clusters to quasi clusters. 
In this article, based on  AQCM, we apply $^{16}$O plus one quasi cluster model for $^{20}$Ne.
Here we focus on the
$E0$ transition matrix element, which has been known as the quantity characterizing the cluster structure.
The $E0$ transition matrix elements are sensitive to the change of the wave functions
from $\alpha$ cluster to $jj$-coupling shell model.
\end{abstract}

\pacs{21.60.Gx, 21.10.Ky}% PACS, the Physics and Astronomy
\maketitle

\section{Introduction}

$^{20}$Ne has been known as a typical example of a nucleus which has $\alpha$
cluster structure. There have been numerous works based on the cluster model, which explain the observed
doublet rotational band structure. 
In addition to the ground $K^\pi = 0^+$ band,
the negative parity band ($K^\pi = 0^-$) starting with the $1^-$ state at
$E_x =  5.787726$ MeV has been observed, 
and existence of this ``low-lying'' negative parity band
is the strong evidence that simple spherical mean field is broken.
These bands are well explained by the picture that
$\alpha$ cluster is located at some distance from the $^{16}$O core \cite{Horiuchi}.
Recently ``container picture'' has been proposed to describe the non-localization of
the $\alpha$ cluster around $^{16}$O \cite{Zhou}.

However, according to the shell model, four nucleons 
perform independent particle motions around the $^{16}$O core, which has doubly 
closed shell of the $p$ shell, and the spin-orbit interaction acts attractively to them.
If we apply simple $\alpha$ cluster models, we cannot take into account this spin-orbit effect.
In traditional $\alpha$ cluster models, $\alpha$ cluster is defined as $(0s)^4$ configuration 
centered at some localized point, and the contributions of non-central interactions vanish. 
If we correctly take into account the spin-orbit effect,
$\alpha$ cluster structure competes with the $jj$-coupling shell model structure.
Previously we have investigated  this competition in $^{20}$Ne
based on the antisymmetrized quasi-cluster model (AQCM) \cite{Ne-Mg}.
AQCM is a method 
that enables us
to describe a transition from the $\alpha$ cluster wave function to the $jj$-coupling shell model
wave function \cite{Simple,Yoshida,Masui,Suhara,General}. 
In this model, the cluster-shell transition is characterized by only two parameters; 
$R$ representing the distance between $\alpha$ cluster and core nucleus 
and $\Lambda$ describing the breaking of the $\alpha$ cluster.
By introducing $\Lambda$, we transform $\alpha$ cluster to quasi cluster, and 
the contribution of the spin-orbit interaction, very important in the 
$jj$-coupling shell model, can be taken into account. 
It was found that the level structure of the yrast states of $^{20}$Ne 
strongly depends on the
strength of the spin-orbit interaction in the Hamiltonian.

In this article we apply AQCM again to $^{20}$Ne and introduce $^{16}$O plus one quasi cluster model.
Particularly we focus on the effect of cluster-shell competition on the $E0$ transition.
The $E0$ transition operator has the form of monopole operator, $\sum_i r^2_i$, and 
this operator changes the nuclear sizes.
However, changing nuclear density uniformly requires quite high excitation energy.
On the other hand,
clusters structures are characterized as weakly interacting states of strongly bound subsystems.
Thus it is rather easy for the cluster states to change the sizes
without giving high excitation energies; this is achieved just by changing the relative distances between clusters.
Therefore, $E0$ transitions in low-energy regions are expected to be signatures of the cluster structures,
and many works along this line are going on 
\cite{Kawabata_PhysLettB646_6,Sasamoto_ModPhysLettA21_2393,Yoshida_PhysRevC79_034308,Yamada_PhysRevC92_034326,Ichikawa_PhysRevC83_061301,Ichikawa_PhysRevC86_031303}.

In our preceding work for $^{16}$O \cite{Matsuno},
we found that
the ground state has a compact four $\alpha$ structure and is almost independent of
the strength of the spin-orbit interaction; however the second $0^+$ state,
which has been known as a $^{12}$C+$\alpha$ cluster state, is very much affected 
by the change of the strength. 
With increasing the strength, the level repulsion and
crossing occur, 
and the $^{12}\mathrm{C}$ cluster part changes from three $\alpha$ 
configuration to the $p_{3/2}$ subclosure of the $jj$-coupling shell model.
The $E0$ transition matrix elements are strongly dependent on this level repulsion and crossing, and they are 
sensitive to the persistence of $4\alpha$ correlation in the excited states.
Here, ``larger cluster'' part of binary cluster system ($^{12}$C part of $^{12}$C+$\alpha$) has been changed
into quasi cluster.
The present study on $^{20}$Ne is different from the preceding work on $^{16}$O
in the following two points. One is that we focus on the change
of ``smaller cluster'' part of the binary cluster system, and in this case, we change 
$\alpha$ cluster around the $^{16}$O core to quasi cluster. 
Another difference is that
this change influences very much the ground state
(in the case of $^{16}$O, the second $0^+$ state with the $^{12}$C+$\alpha$ configuration
is affected by the spin-orbit interaction). 
Since other higher nodal states are
determined by the orthogonal condition to the ground state, this change also has influences
on the
wave functions of the excited states. Naturally  $E0$ transition matrix elements
are also affected by this change.

The paper is organized as follows. 
The formulation is given in Sect.~\ref{model}. 
In Sect.~\ref{results}, the results for $^{20}$Ne are shown. 
Finally, in Sect.~\ref{summary} we summarize the results and give the main conclusion.

\section{Formulation}\label{model}

\subsection{Wave function of the total system}

The wave function of the total system $\Psi$ is antisymmetrized product of these
single particle wave functions;
\begin{equation}
\Psi = {\cal A} \{ (\psi_1 \chi_1 \tau_1) (\psi_2 \chi_2 \tau_2) (\psi_3 \chi_3 \tau_3) \cdot \cdot \cdot \cdot (\psi_A \chi_A \tau_A)\}.
\label{total-wf}
\end{equation}  
The projection onto parity and angular momentum eigen states can be numerically performed.
The number of mesh points for the integral over Euler angles is $16^3$.

\subsection{Single particle orbits -- $^{16}$O part}

For the single particle orbits of the $^{16}$O part,
we introduce conventional $\alpha$ cluster model. 
The single particle wave function has a Gaussian shape \cite{Brink};
\begin{equation}
	\phi_{i} = \left( \frac{2\nu}{\pi} \right)^{\frac{3}{4}}
	\exp \left[- \nu \left(\bm{r}_{i} - \bm{R}_i \right)^{2} \right] \eta_{i},
\label{Brink-wf}
\end{equation}
where $\eta_{i}$ represents the spin-isospin part of the wave function, 
and $\bm{R}_i$ is a real parameter representing the center of a Gaussian 
wave function for the $i$th particle. 
For the width parameter, we use the value of $b = 1.6$ fm, $\nu = 1/2b^2$.
In this Brink-Bloch wave function, four nucleons in one $\alpha$ cluster share the common $\bm{R}_i$ value. 
Hence, the contribution of the spin-orbit interaction vanishes. 
We introduce four different kinds of $\bm{R}_i$ values, and four $\alpha$ clusters 
are forming tetrahedron configuration.
When we take the limit of the relative distances between $\alpha$ cluster to zero,
the wave function coincide with the closed $p$ shell configuration of the shell model \cite{Brink},
and this limit is called Elliot SU(3) limit \cite{Elliot}.
In our model, the relative distance is taken to be a small value, 0.1 fm.

\subsection{Single particle orbits -- one quasi cluster part}

We add one quasi cluster around the $^{16}$O core based on AQCM.
In the AQCM, $\alpha$ clusters are changed into quasi clusters. 
For nucleons in the quasi cluster, 
the single particle wave function is described by 
a Gaussian wave packet, and
the center of this packet $\bm{\zeta}_{i}$ is a complex parameter;
\begin{equation}
	\psi_{i} = \left( \frac{2\nu}{\pi} \right)^{\frac{3}{4}}
		\exp \left[- \nu \left(\bm{r}_{i} - \bm{\zeta}_{i} \right)^{2} \right] \chi_{i} \tau_{i}, 
\label{AQCM_sp} 
\end{equation}
\begin{equation}
	\bm{\zeta}_{i} = \bm{R}_i + i \Lambda \bm{e}^{\text{spin}}_{i} \times \bm{R}_i, 
\label{condition}
\end{equation}
where
$\chi_{i}$ and $\tau_{i}$ in Eq.~\eqref{AQCM_sp} represent the intrinsic spin and isospin part of the $i$th 
single particle wave function, respectively. In Eq~\eqref{condition},
$\bm{e}^{\text{spin}}_{i}$ 
is a unit vector for the orientation of the intrinsic spin $\chi_{i}$.
Here, $\Lambda$ is a real control parameter describing the dissolution of the $\alpha$ cluster. 
The width parameter is the same as nucleons in the $^{16}$O cluster ($b = 1.6$ fm, $\nu = 1/2b^2$). 
As one can see immediately, the $\Lambda = 0$ AQCM wave function, which has no imaginary part, 
is the same as the conventional Brink-Bloch wave function.
The AQCM wave function corresponds to the $jj$-coupling shell model wave function
when $\Lambda = 1$ and $\bm{R}_i \rightarrow 0$.
The mathematical explanation is summarized
in Ref.~\cite{General}.

Gaussian center parameters for the four nucleons in the quasi cluster 
($\zeta_{17} \sim \zeta_{20}$) are given in the following way.
Firstly.
we place quasi cluster on the $z$ axis, 
and the real part of the Gaussian center parameters 
($\bm{R_{17 \sim 20}}$)
are given as
\begin{equation}
\bm{R_{17}} = \bm{R_{18}} = \bm{R_{19}} = \bm{R_{20}} = R\bm{e_z}.
\end{equation}
Here
$R$ is a parameter, which describes the distance between quasi cluster and the $^{16}$O cluster,
and $\bm{e_z}$ is the unit vector in the $z$ direction.
Next, we give the imaginary parts.
Here we quantize the spin of the nucleons
along the $x$ axis, and 
in order to satisfy the condition of Eq.~\eqref{condition},
we must give the imaginary parts in the $-y$ direction as,
\begin{equation}
\bm{\zeta}_{17} = R(\bm{e_z} - i \Lambda \bm{e_y}),
\end{equation}
\begin{equation}
\bm{\zeta}_{18} = R(\bm{e_z} + i \Lambda \bm{e_y}),
\end{equation}
\begin{equation}
\bm{\zeta}_{19} = R(\bm{e_z} - i \Lambda \bm{e_y}),
\end{equation}
\begin{equation}
\bm{\zeta}_{20} = R(\bm{e_z} + i \Lambda \bm{e_y}),
\end{equation}
where $e_z$ and $e_y$ are unit vectors in the $z$ and $y$ direction, respectively.
The Gaussian center parameter 
$\bm{\zeta}_{17}$ is for a proton with spin up ($+x$ direction),
$\bm{\zeta}_{18}$ is for a proton with spin down ($-x$ direction),
$\bm{\zeta}_{19}$ is for a neutron with spin up ($+x$ direction),
and
$\bm{\zeta}_{20}$ is for a neutron with spin down ($-x$ direction).
When $\Lambda$ is set to zero, the wave function consisting the quasi clusters
agrees with that of an  $\alpha$ cluster. If we take the limit of $R \to 0$ and $\Lambda =1$,
four nucleons in the quasi cluster occupy $d_{5/2}$ orbits of the $jj$-coupling shell model.

\subsection{Hamiltonian}
For the Hamiltonian, 
we use
 Volkov No.2 \cite{Vol} as an effective interaction 
for the central part with the Majorana exchange parameter of $M = 0.62$.
For the spin-orbit part,
G3RS \cite{G3RS}, which is a realistic
interaction originally determined to reproduce the nucleon-nucleon scattering phase shift, 
is adopted;
\begin{equation}
\hat{V}_{spin-orbit}= V_{ls}( e^{-d_{1}r^{2}}-e^{-d_{2}r^{2}}) P(^{3}O){\vec{L}}\cdot{\vec{S}},
\end{equation}
 where $d_{1}= 5. 0$ fm$^{-2},\ d_{2}= 2. 778$ fm$^{-2}$,
and $P(^{3}O)$ is a projection operator onto a triplet odd state.
The operator $\vec{L}$ stands for the relative angular momentum
 and $\vec{S}$ is the spin ($\vec{S_{1}}+\vec{S_{2}}$).
In the present work, the strength of the spin-orbit interaction, $V_{ls}$, is a parameter as in Ref. \cite{Matsuno} 
and we compare the results by changing the value.

\section{Results}\label{results}

In this section, we apply our AQCM wave function introduced in the
previous section to $^{20}$Ne and discuss the $V_{ls}$ 
(strength of the spin-orbit interaction) dependence
of energy levels and $E0$ transition probabilities.
The $V_{ls}$ value is changed from 0 MeV to 3000 MeV,
and reasonable value of  around $1500$ MeV has been suggested
in our preceding work \cite{Ne-Mg}.

\subsection{Energy of each GCM basis state}

We prepare AQCM wave functions 
with different $R$ and $\Lambda$ values
as basis states of
generator coordinate method (GCM).
The adopted
values are $R = 1,2,3,4,5,6,7$ fm and $\Lambda = 0, 1/3, 2/3, 1$.
The $0^+$ energies of these basis states are presented 
in Table~\ref{basis}, and here, we show the values for two extreme 
cases for the 
strengths of the spin-orbit interaction;
(a) $V_{ls} = 0$ MeV and (b) $V_{ls} = 3000$ MeV.
The $0^+$ energies of the GCM basis states corresponding to other $V_{ls}$
values can be estimated just by interpolating these values linearly.

In Table~\ref{basis}~(a), we find that $\Lambda = 0$ basis states
give lower energies than $\Lambda$ finite basis states.
This is because of the absence of the spin-orbit interaction;
introducing imaginary part for the Gaussian center parameters
does not work for the spin-orbit interaction and that
simply increases the kinetic energy of four nucleons in the quasi cluster. 
Here the basis state with $R = 3$ fm ($\Lambda = 0$) gives the lowest energy
of $-153.6$ MeV.

On the contrary, Table~\ref{basis}~(b) is the case of  $V_{ls} = 3000$ MeV,
and basis states with finite $\Lambda$ values get much lower,
since the contribution of the spin-orbit interaction can be taken into account
by transforming the $\alpha$ cluster to quasi cluster.
The basis state which gives the lowest energy has the values of $\Lambda = 2/3$ and $R = 1$ fm
($-163.8$ MeV).
This result suggests that when
the spin-orbit interaction is switched on,
the $R$ value of the optimal basis state becomes smaller and
the $\Lambda$ value increases. This means that not only the $\alpha$ cluster dissolutes into 
quasi cluster, the relative distance between the cluster and the $^{16}$O core decreases.

\begin{table} 
 \caption{The $0^+$ energies of GCM basis states for the cases of
different strengths of the spin-orbit interaction; 
(a) $V_{ls} = 0$ MeV and (b)  $V_{ls} = 3000$ MeV.
The $0^+$ energies of the GCM basis states with other $V_{ls}$
values can be estimated by linearly interpolating these two.
 }

(a)

  \begin{tabular}{ccccc} 
\hline 
\hline 
   $R$ (fm)  &  $\Lambda=0$ & $\Lambda=1/3$ & $\Lambda=2/3$ & $\Lambda=1$ \\ 
\hline
1   &  $-147.7$  &  $-141.8$ & $-131.5$  & $-128.1$ \\  
2   &  $-151.5$  &  $-144.1$ & $-127.5$  & $-111.7$ \\  
3   &  $-153.6$  &  $-141.2$ & $-104.6$  & $-50.0$ \\  
4   &  $-152.1$  &  $-129.4$ & $-55.3$   & $54.5$ \\  
5   &  $-148.2$  &  $-109.1$ & $8.7$      & $183.4$ \\  
6   &  $-144.5$  &  $-84.3$  &  $77.7$    & $330.4$ \\  
7   &  $-142.7$  &  $-59.1$  &  $151.8$  & $489.9$ \\  
\hline
\end{tabular} 

(b)

  \begin{tabular}{ccccc} 
\hline 
\hline 
   $R$ (fm)  &  $\Lambda=0$ & $\Lambda=1/3$ & $\Lambda=2/3$ & $\Lambda=1$ \\ 
\hline
1   &  $-147.7$  &  $-163.8$ & $-164.2$  & $-162.4$ \\  
2   &  $-151.5$  &  $-162.3$ & $-155.9$  & $-144.3$ \\  
3   &  $-153.6$  &  $-153.3$ & $-124.6$  & $-76.6$ \\  
4   &  $-152.1$  &  $-134.4$ & $-63.6$   & $41.3$ \\  
5   &  $-148.2$  &  $-108.3$ & $9.9$      & $182.6$ \\  
6   &  $-144.5$  &  $-80.2$  &  $83.0$    & $334.5$ \\  
7   &  $-142.7$  &  $-53.6$  &  $158.8$  & $503.5$ \\  
\hline
  \end{tabular}       

\label{basis}
\end{table}

\subsection{$0^+$ energies}

We superpose these AQCM wave functions with different $R$ and $\Lambda$ values
based on GCM. The $0^+$ eigen energies and coefficients for the linear combination of the
GCM basis states for each eigen state are obtained by diagonalizing the Hamiltonian
(solving the Hill-Wheeler equation \cite{Brink}). 
Here we change the strength of the spin-orbit interaction, $V_{ls}$, and 
diagonalize the Hamiltonian at each $V_{ls}$ value.

\begin{figure}[t]
	\centering
	\includegraphics[width=6.0cm]{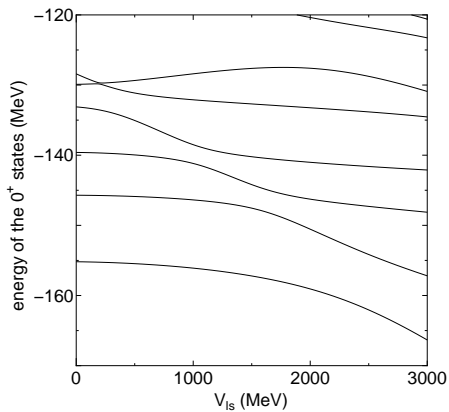} 
	\caption{
The $0^+$ energy curves of $^{20}$Ne obtained by superposing
the AQCM wave functions with $R=1,2,3,4,5,6,7$ fm and $\Lambda = 0, 1/3, 2/3, 1$.
The energy curves are plotted as a function of
the strength of the spin-orbit interaction, $V_{ls}$ in the Hamiltonian
     }
\label{vls-dep.0+}
\end{figure}

The obtained $0^+$ energy curves of $^{20}$Ne as a function of
$V_{ls}$ are shown in Fig.~\ref{vls-dep.0+}.
The ground state gets more binding with increasing $V_{ls}$;
the $0^+$ energy changes from $-155.2$ MeV ($V_{ls} = 0$ MeV)
to $-166.3$ MeV ($V_{ls} = 3000$ MeV). 
These energies are lower than the ones for the optimal GCM basis states
shown in Table~\ref{basis} by about 2 MeV, and this is the effect of superposing
the GCM basis states. 
The experimental value
for the ground state is $-160.6448$ MeV.

In Fig.~\ref{vls-dep.0+}, we find that the fourth $0^+$ state at $V_{ls} = 0$ starts lowering soon after the spin-orbit interaction
is switched on, and the decrease of the energy is much steeper than other states.
The level repulsion (crossing) between the fourth and the third $0^+$ states occurs around $V_{ls} = 1000$ MeV,
and here the wave functions of these two states strongly mix.
There is another level repulsion (crossing) between this third and the second $0^+$ state around $V_{ls} = 1500$ MeV.
Because of these level repulsions (crossings), it looks that 
the wave function of the fourth state at $V_{ls} = 0$ MeV 
comes down and mixes in the ground and second states at $V_{ls} = 3000$ MeV. 

On the other hand, the wave function of the second  $0^+$
state at $V_{ls} = 0$ MeV almost stays at this
energy even after the spin-orbit interaction is switched on. 
Around $V_{ls} = 1500$ MeV,
the level repulsion (crossing) occurs and
this wave function  becomes the one for the third $0^+$ state beyond this region, but the energy is almost constant even after that. 
Similar thing can be found for the third $0^+$ state at $V_{ls} = 0$.
After the level repulsion (crossing) around $V_{ls} = 1000$ MeV,
this state corresponds to the fourth $0^+$ state at $V_{ls} = 3000$ MeV.
These two states are considered to return back to the $\alpha$ cluster structure
beyond these level repulsion (crossing) regions.

\subsection{Intrinsic spin}

\begin{figure}[t]
	\centering
	\includegraphics[width=6.0cm]{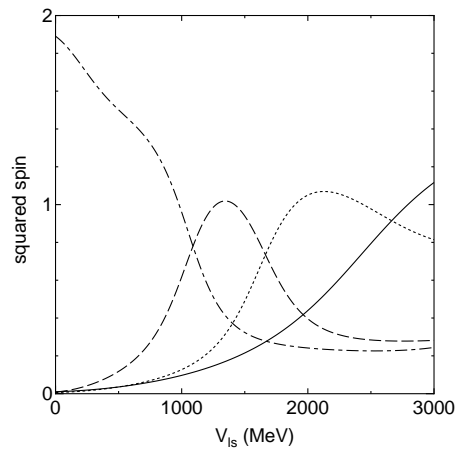} 
	\caption{
The absolute value of the expectation values for the squared spin as a function
of the strength of the spin-orbit interaction, $V_{ls}$.
The solid, dotted, dashed, and dash-dotted lines are for the
ground, second, third, and fourth $0^+$ states of Fig.~\ref{vls-dep.0+}.
     }
\label{vls-dep.ssq}
\end{figure}

Next, we discuss the structure change of each $0^+$ state
as a function of  $V_{ls}$ by analyzing the spin structure. 
In the traditional cluster models, such as $^{16}$O+$\alpha$ models, 
the clusters are spin saturated systems, and the expectation value of intrinsic spin operator
($\sum_i \vec s_i$, where $\vec s_i$ is the spin operator for the $i$th nucleon) becomes zero.
However, in AQCM, $\alpha$ clusters are changed into quasi clusters,
and the contribution of the spin-orbit interaction can be taken into account.
In such case, the intrinsic spin structure of quasi cluster changes from
that of $\alpha$ cluster as a function of $\Lambda$ value.
This can be proven by calculating the  expectation values for the square of the intrinsic spin operator 
($\sum_{i,j} \vec s_i \cdot \vec s_j$).

In Fig.~\ref{vls-dep.ssq}, the absolute values of the expectation value
for the squared spin are shown
as a function
of the strength of the spin-orbit interaction, $V_{ls}$.
The solid, dotted, dashed, and dash-dotted lines are for the
ground, second, third, and fourth $0^+$ states of Fig.~\ref{vls-dep.0+}.
At $V_{ls} = 0$ MeV, the ground, second, and third $0^+$ states have squared spin zero;
without the spin-orbit force, $\alpha$ cluster structure is not broken.
On the contrary, the fourth $0^+$ state has the value of 1.89.
Here the $\alpha$ cluster structure is broken even without the spin-orbit interaction,
and this is considered to be 
due to orthogonal condition to other lower states.

With increasing $V_{ls}$, the values for the first, second, and third $0^+$ states start increasing.
This corresponds to the fact that the spin-orbit interaction acts attractively for these states.
Around $V_{ls} = 1000$ MeV, the dashed line and dash-dotted line cross,
and this is due to the level repulsion (crossing) of the third and fourth states
shown in Fig.~\ref{vls-dep.0+}.
Also, this dashed line crosses with dotted line around 
$V_{ls} = 1500$ MeV, and this corresponds to the level repulsion (crossing) of the
second and third states, as discussed in the previous subsection.
Beyond this level repulsion region, the values for the third and fourth $0^+$ states
decrease, and $\alpha$ cluster components become important again in these states.
The third and fourth $0^+$ states go back to $\alpha$ cluster structure.
On the contrary, the values for the first and second $0^+$ states 
significantly increase around $V_{ls} = 2000$ MeV, and in this region, 
it is considered that
the component of fourth $0^+$ state at $V_{ls} = 0$ MeV strongly mix
in these states.

\subsection{Squared overlap between the final solution and each GCM basis state}

\begin{table} 
 \caption{The absolute values of the squared overlaps 
between the final solution and each GCM basis state in the case of $V_{ls} = 0$ MeV.
(a) is for the ground $0^+$ state ($-155.2$ MeV),
(b) is for the second $0^+$ state ($-145.7$ MeV), 
(c) is for the third $0^+$ state ($-139.6$ MeV), 
and (d) is for the fourth $0^+$ state ($-133.1$ MeV).
 }

(a)

  \begin{tabular}{ccccc} 
\hline 
\hline 
   $R$ (fm)  &  $\Lambda=0$ & $\Lambda=1/3$ & $\Lambda=2/3$ & $\Lambda=1$ \\ 
\hline
1   &  0.57  &  0.35 & 0.07  & 0.03 \\  
2   &  0.74  &  0.45 & 0.08  & 0.02 \\  
3   &  0.92  &  0.53 & 0.07  & 0.01 \\  
4   &  0.83  &  0.39 & 0.02  & 0.00 \\  
5   &  0.43  &  0.13 & 0.00  & 0.00 \\  
6   &  0.11  &  0.02 & 0.00  & 0.00 \\  
7   &  0.02  &  0.00 & 0.00  & 0.00 \\  
\hline
\end{tabular} 

(b)

  \begin{tabular}{ccccc} 
\hline 
\hline 
   $R$ (fm)  &  $\Lambda=0$ & $\Lambda=1/3$ & $\Lambda=2/3$ & $\Lambda=1$ \\ 
\hline
1   &  0.17  &  0.12 & 0.03  & 0.01 \\  
2   &  0.15  &  0.09 & 0.02  & 0.00 \\  
3   &  0.04  &  0.03 & 0.01  & 0.00 \\  
4   &  0.03  &  0.01 & 0.00  & 0.00 \\  
5   &  0.41  &  0.11 & 0.00  & 0.00 \\  
6   &  0.74  &  0.11 & 0.00  & 0.00 \\  
7   &  0.53  &  0.03 & 0.00  & 0.00 \\  
\hline
  \end{tabular}    

(c)

  \begin{tabular}{ccccc} 
\hline 
\hline 
   $R$ (fm)  &  $\Lambda=0$ & $\Lambda=1/3$ & $\Lambda=2/3$ & $\Lambda=1$ \\ 
\hline
1   &  0.14  &  0.10 & 0.03  & 0.21 \\  
2   &  0.08  &  0.06 & 0.02  & 0.01 \\  
3   &  0.00  &  0.01 & 0.00  & 0.00 \\  
4   &  0.08  &  0.03 & 0.00  & 0.00 \\  
5   &  0.12  &  0.03 & 0.00  & 0.00 \\  
6   &  0.02  &  0.00 & 0.00  & 0.00 \\  
7   &  0.43  &  0.02 & 0.00  & 0.00 \\  
\hline
  \end{tabular}    

(d)

  \begin{tabular}{ccccc} 
\hline 
\hline 
   $R$ (fm)  &  $\Lambda=0$ & $\Lambda=1/3$ & $\Lambda=2/3$ & $\Lambda=1$ \\ 
\hline 
1   &  0.00  &  0.32 & 0.33  & 0.23 \\  
2   &  0.00  &  0.34 & 0.32  & 0.16 \\  
3   &  0.00  &  0.27 & 0.17  & 0.03 \\  
4   &  0.00  &  0.12 & 0.03  & 0.00 \\  
5   &  0.00  &  0.02 & 0.00  & 0.00 \\  
6   &  0.00  &  0.00 & 0.00  & 0.00 \\  
7   &  0.00  &  0.00 & 0.00  & 0.00 \\  
\hline
  \end{tabular}    

\label{sqol-0MeV}
\end{table}

Next we discuss the character of each 
state by showing the squared overlap between the final solution 
and each GCM basis state.
In Table~\ref{sqol-0MeV}, the absolute values of the squared overlaps 
between the final solution and each GCM basis state in the case of $V_{ls} = 0$ MeV are shown.
These are the results when the spin-orbit interaction is switched off, and
(a) is for the ground $0^+$ state ($-155.2$ MeV),
(b) is for the second $0^+$ state ($-145.7$ MeV), 
(c) is for the third $0^+$ state ($-139.6$ MeV), 
and (d) is for the fourth $0^+$ state ($-133.1$ MeV).
As shown in Table~\ref{sqol-0MeV}~(a),
the ground $0^+$ state has the squared overlap of 0.92
with the GCM basis state which has $R = 3$ fm and $\Lambda =0$. 
The second $0^+$ state is a higher nodal state 
and has much larger $^{16}$O-$\alpha$ distance than the ground state;
this is due to the  orthogonal condition to the ground state.
As shown in Table~\ref{sqol-0MeV}~(b), 
the state has the squared overlap of 0.74 (0.53)
with the GCM basis state which has $R = 6$ (7) fm and $\Lambda =0$.
The third $0^+$ state also has large $^{16}$O-$\alpha$ distance (Talbe~\ref{sqol-0MeV}~(c)).
The fourth $0^+$ state has overlaps with basis states with finite $\Lambda$ values
(Table~\ref{sqol-0MeV}~(b)). However, since the spin-orbit interaction
is absent ($V_{ls} = 0$ MeV), the excitation energy is rather large ($E_x \sim $ 22.1 MeV).

\begin{table} 
 \caption{The absolute values of the squared overlaps 
between the final solution and each GCM basis state in the case of $V_{ls} = 3000$ MeV.
(a) is for the ground $0^+$ state ($-166.3$ MeV),
(b) is for the second $0^+$ state ($-157.2$ MeV), 
(c) is for the third $0^+$ state ($-148.1$ MeV), 
and (d) is for the fourth $0^+$ state ($-142.1$ MeV).
 }

(a)

  \begin{tabular}{ccccc} 
\hline 
\hline 
   $R$ (fm)  &  $\Lambda=0$ & $\Lambda=1/3$ & $\Lambda=2/3$ & $\Lambda=1$ \\ 
\hline
1   &  0.33  &  0.84 & 0.88  & 0.74 \\  
2   &  0.32  &  0.79 & 0.74  & 0.49 \\  
3   &  0.27  &  0.53 & 0.32  & 0.07 \\  
4   &  0.13  &  0.18 & 0.00  & 0.00 \\  
5   &  0.03  &  0.02 & 0.00  & 0.00 \\  
6   &  0.00  &  0.00 & 0.00  & 0.00 \\  
7   &  0.00  &  0.00 & 0.00  & 0.00 \\  
\hline
\end{tabular} 

(b)

  \begin{tabular}{ccccc} 
\hline 
\hline 
   $R$ (fm)  &  $\Lambda=0$ & $\Lambda=1/3$ & $\Lambda=2/3$ & $\Lambda=1$ \\ 
\hline
1   &  0.32  &  0.08 & 0.10  & 0.21 \\  
2   &  0.42  &  0.14 & 0.05  & 0.11 \\  
3   &  0.49  &  0.21 & 0.00  & 0.00 \\  
4   &  0.40  &  0.17 & 0.00  & 0.00 \\  
5   &  0.18  &  0.06 & 0.00  & 0.00 \\  
6   &  0.04  &  0.01 & 0.00  & 0.00 \\  
7   &  0.00  &  0.00 & 0.00  & 0.00 \\  
\hline
  \end{tabular}    

(c)

  \begin{tabular}{ccccc} 
\hline 
\hline 
   $R$ (fm)  &  $\Lambda=0$ & $\Lambda=1/3$ & $\Lambda=2/3$ & $\Lambda=1$ \\ 
\hline
1   &  0.04  &  0.05 & 0.00  & 0.01 \\  
2   &  0.01  &  0.02 & 0.00  & 0.01 \\  
3   &  0.03  &  0.01 & 0.01  & 0.00 \\  
4   &  0.32  &  0.13 & 0.01  & 0.00 \\  
5   &  0.67  &  0.20 & 0.00  & 0.00 \\  
6   &  0.57  &  0.09 & 0.00  & 0.00 \\  
7   &  0.24  &  0.01 & 0.00  & 0.00 \\  
\hline
  \end{tabular} 

(d)

  \begin{tabular}{ccccc}
\hline 
\hline 
   $R$ (fm)  &  $\Lambda=0$ & $\Lambda=1/3$ & $\Lambda=2/3$ & $\Lambda=1$ \\ 
\hline
1   &  0.00  &  0.01 & 0.00  & 0.00 \\  
2   &  0.01  &  0.00 & 0.00  & 0.00 \\  
3   &  0.06  &  0.02 & 0.01  & 0.00 \\  
4   &  0.10  &  0.05 & 0.00  & 0.00 \\  
5   &  0.00  &  0.00 & 0.00  & 0.00 \\  
6   &  0.26  &  0.03 & 0.00  & 0.00 \\  
7   &  0.67  &  0.04 & 0.00  & 0.00 \\  
\hline
  \end{tabular} 

\label{sqol-3000MeV}
\end{table}

We move on to another extreme case that the strength of the spin-orbit interaction is set to
${V_{ls} = 3000}$ MeV.  
In Table~\ref{sqol-3000MeV}, the absolute values of the squared overlaps 
between the final solution and each GCM basis state in the case of $V_{ls} = 3000$ MeV
are shown. Here
(a) is for the ground $0^+$ state ($-166.3$ MeV),
(b) is for the second $0^+$ state ($-157.2$ MeV), 
(c) is for the third $0^+$ state ($-148.1$ MeV), 
and (d) is for the fourth $0^+$ state ($-142.1$ MeV).
As shown in Table~\ref{sqol-3000MeV}~(a),
the ground $0^+$ state has the squared overlap of 0.88
with the GCM basis state which has $R = 1$ fm and $\Lambda =2/3$.
The $R$ value becomes very small and $\Lambda$ value increased 
compared with the case of $V_{ls} = 0$ MeV, as expected in the previous subsection. 
The $\alpha$ cluster structure is completely washed out.
The second $0^+$ state is no longer a higher nodal state of $^{16}$O+$\alpha$
with large relative distance, because of the level crossing when increasing the $V_{ls}$ value. 
As shown in Table~\ref{sqol-3000MeV}~(b), 
although the state still has the squared overlap of 0.49
with the GCM basis state which has $R = 3$ fm and $\Lambda =0$,
the squared overlaps with finite $\Lambda$ basis states increase.
The third and fourth $0^+$ states have overlaps with basis states with large $R$ values;
the third state has 0.67 with $R = 5$ fm $\Lambda = 0$ (Table~\ref{sqol-3000MeV}~(c)),
and the forth state has 0.67 
with $R = 7$ fm $\Lambda = 0$ (Table~\ref{sqol-3000MeV}~(d)).
The character of the second and third $0^+$ states at $V_{ls} = 0$ MeV remains here,
as expected in the previous subsection.

\subsection{$E0$ transition matrix elements}

\begin{figure}[t]
	\centering
	\includegraphics[width=6.0cm]{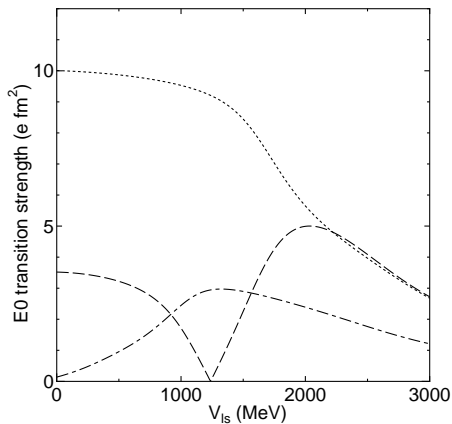} 
	\caption{
The $E0$ transition matrix elements from the ground state.
The dotted, dashed, and dash-dotted lines show the ones
to the second, third, and fourth $0^+$ states.
     }
\label{e0}
\end{figure}

The $E0$ transition matrix elements from the ground state are shown in Fig.~\ref{e0}.
The dotted, dashed, and dash-dotted lines show the ones
to the second, third, and fourth $0^+$ states.
The operator has the form of $e\sum_i r_i^2$, and summation is for protons.
The experimental value form the ground state to the second $0^+$ state is 6.914 $e$ fm$^2$.
The calculated value (dotted line) at $V_{ls} = 0$ MeV is  10.0 $e$ fm$^2$, and 
this is slightly larger.
As we discussed in Table~\ref{sqol-0MeV} (a) and (b),
when the spin-orbit interaction is switched off, both
ground and second $0^+$ states have $\alpha$ cluster structure,
and the second $0^+$ state has spatially more extended distribution.
In this case the matrix element of the $E0$ transition between these states becomes large.
With increasing the $V_{ls}$ value, the mixing of basis states with finite $\Lambda$ values
becomes important in both the ground and second $0^+$ states,
and the $E0$ transition matrix decreases.
The value agrees with the experimental one around $V_{ls} = 1770$ MeV.
In our preceding work \cite{Ne-Mg}, we deduced proper strength for the spin-orbit interaction
from the analyses on the level structure, and the present result is almost similar to this one.

As shown in Fig.~\ref{e0},
around $V_{ls} = 1000 \sim 1500$ MeV region, the dashed line and dash-dotted line cross,
reflecting the fact that the level repulsion (crossing) of the third and fourth states occurs.
The wave functions are interchanged in this region.
Also, this dashed line crosses with the dotted line around 
$V_{ls} = 2000$ MeV, and this corresponds to the level repulsion (crossing) between the
third and second and states  
as discussed in the previous subsection.

\begin{table} 
 \caption{The absolute values of the squared overlaps 
between the final solution and each GCM basis state in the case of $V_{ls} = 1770$ MeV,
which gives a reasonable $E0$ transition matrix element from the ground state to the second $0^+$ state.
(a) is for the ground $0^+$ state ($-158.1$ MeV),
(b) is for the second $0^+$ state ($-148.9$ MeV), 
(c) is for the third $0^+$ state ($-145.5$ MeV), 
and (d) is for the fourth $0^+$ state ($-140.7$ MeV).
 }

(a)

  \begin{tabular}{ccccc} 
\hline 
\hline 
   $R$ (fm)  &  $\Lambda=0$ & $\Lambda=1/3$ & $\Lambda=2/3$ & $\Lambda=1$ \\ 
\hline
1   &  0.62  &  0.74 & 0.33  & 0.20 \\  
2   &  0.73  &  0.82 & 0.33  & 0.15 \\  
3   &  0.78  &  0.75 & 0.19  & 0.03 \\  
4   &  0.57  &  0.40 & 0.03  & 0.00 \\  
5   &  0.23  &  0.09 & 0.00  & 0.00 \\  
6   &  0.05  &  0.01 & 0.00  & 0.00 \\  
7   &  0.01  &  0.00 & 0.00  & 0.00 \\  
\hline
\end{tabular} 

(b)

  \begin{tabular}{ccccc} 
\hline 
\hline 
   $R$ (fm)  &  $\Lambda=0$ & $\Lambda=1/3$ & $\Lambda=2/3$ & $\Lambda=1$ \\ 
\hline
1   &  0.00  &  0.18 & 0.49  & 0.50 \\  
2   &  0.00  &  0.11 & 0.38  & 0.32 \\  
3   &  0.07  &  0.01 & 0.12  & 0.04 \\  
4   &  0.26  &  0.04 & 0.00  & 0.00 \\  
5   &  0.40  &  0.09 & 0.00  & 0.00 \\  
6   &  0.27  &  0.04 & 0.00  & 0.00 \\  
7   &  0.10  &  0.01 & 0.00  & 0.00 \\  
\hline
  \end{tabular}    

(c)

  \begin{tabular}{ccccc} 
\hline 
\hline 
   $R$ (fm)  &  $\Lambda=0$ & $\Lambda=1/3$ & $\Lambda=2/3$ & $\Lambda=1$ \\ 
\hline
1   &  0.11  &  0.00 & 0.15  & 0.23 \\  
2   &  0.10  &  0.00 & 0.12  & 0.14 \\  
3   &  0.04  &  0.00 & 0.05  & 0.02 \\  
4   &  0.01  &  0.02 & 0.01  & 0.00 \\  
5   &  0.24  &  0.08 & 0.00  & 0.00 \\  
6   &  0.47  &  0.07 & 0.00  & 0.00 \\  
7   &  0.36  &  0.02 & 0.00  & 0.00 \\  
\hline
  \end{tabular} 

(d)

  \begin{tabular}{ccccc}
\hline 
\hline 
   $R$ (fm)  &  $\Lambda=0$ & $\Lambda=1/3$ & $\Lambda=2/3$ & $\Lambda=1$ \\ 
\hline
1   &  0.05  &  0.04 & 0.00  & 0.02 \\  
2   &  0.01  &  0.01 & 0.01  & 0.01 \\  
3   &  0.01  &  0.00 & 0.01  & 0.00 \\  
4   &  0.11  &  0.05 & 0.00  & 0.00 \\  
5   &  0.06  &  0.02 & 0.00  & 0.00 \\  
6   &  0.08  &  0.01 & 0.00  & 0.00 \\  
7   &  0.49  &  0.03 & 0.00  & 0.00 \\  
\hline
  \end{tabular} 

\label{sqol-1770MeV}
\end{table}

Now we analyze the wave function of each $0^+$ state calculated using the spin-orbit strength
which reproduces the experimental $E0$ transition probability 
from the ground state to the second $0^+$ state.
The absolute value of the squared overlap 
between the final solution and each GCM basis state in the case of $V_{ls} = 1770$ MeV is
shown in Table~\ref{sqol-1770MeV};
(a) is for the ground $0^+$ state ($-158.1$ MeV),
(b) is for the second $0^+$ state ($-148.9$ MeV), 
(c) is for the third $0^+$ state ($-145.5$ MeV), 
and (d) is for the fourth $0^+$ state ($-140.7$ MeV).
The ground state has the squared overlap with the $\alpha$ cluster state;
the value for the basis state with $R = 3$ fm and $\Lambda =0$ 
is 0.78, and the character at $V_{ls} = 0$ MeV still remains.
However the largest squared overlap of 0.82 is with the basis state which has
$R = 2$ fm and $\Lambda = 1/3$. Therefore, the $\alpha$ breaking effect
due to the spin-orbit interaction is important.
The second $0^+$ states was very extended $^{16}$O+$\alpha$ cluster state
at $V_{ls} = 0$ MeV, and the second state at $V_{ls} = 1770$ MeV
still has the squared overlap of 0.40 with the basis state which has $R = 5$ fm and $\Lambda = 0$.
However, 
at  $V_{ls} = 1770$ MeV the state also
has components of the basis states with finite $\Lambda$ values;
the squared overlap with the basis state $R = 1$ fm and $\Lambda =1$ is 0.50.
The third and fourth $0^+$ states at $V_{ls} = 1770$ MeV contain the components of
$\alpha$ cluster structure ($\Lambda =0$)  with large $R$ values.

\subsection{$1^-$ states}

\begin{figure}[t]
	\centering
	\includegraphics[width=6.0cm]{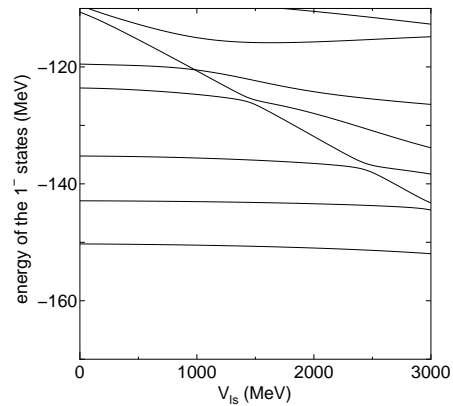} 
	\caption{
The $1^-$ energy curves of $^{20}$Ne obtained by superposing
the AQCM wave functions with $R=1,2,3,4,5,6,7$ fm and $\Lambda = 0, 1/3, 2/3, 1$.
The energy curves are plotted as a function of
the strength of the spin-orbit interaction, $V_{ls}$ in the Hamiltonian
     }
\label{vls-dep.1-}
\end{figure}

We move on from $0^+$ states to $1^-$ states,
and the energy curves of $1^-$ states ($K=0$)
as a function of $V_{ls}$
are shown in Fig.~\ref{vls-dep.1-}.
The presence of low-lying negative parity band starting with the first $1^-$
has been the key evidence for the $\alpha$ cluster structure.
The present result shows that the energy of this first $1^-$ state 
is almost constant even if the spin-orbit interaction is switched on.
This means that $\alpha$ breaking basis states do not contribute 
and the $\alpha$ cluster structure is really important in this state.
The absolute values of the squared overlaps 
between the first $1^-$ state and each GCM basis state
are shown in Table~\ref{sqol-1-}.
Table~\ref{sqol-1-}~(a) is the case of $V_{ls} = 0$ MeV, which gives $-150.3$ MeV
for the first $1^-$ state.
The largest squared overlap of 0.89 is 
with the base state which has $R = 4$ fm and $\Lambda =$ 0.
This character remains in Table~\ref{sqol-1-}~(b), which is the case of $V_{ls} = 3000$ MeV. 
The first $1^-$ state is obtained at $-152.0$ MeV,
and the largest squared overlap of 0.85 is 
with the base state which has $R = 4$ fm and $\Lambda =$ 0.
Even in the case of quite strong spin-orbit interaction, the $\alpha$
cluster structure remains in the first $1^-$ state.

\begin{table} 
 \caption{The absolute values of the squared overlaps 
between the first $1^-$ state and each GCM basis state.
(a) is at $V_{ls} = 0$ MeV and $1^-_1$ is obtained at $-150.3$ MeV.
(b) is at $V_{ls} = 3000$ MeV and $1^-_1$ is obtained at $-152.0$ MeV..
 }

(a)

  \begin{tabular}{ccccc} 
\hline 
\hline 
   $R$ (fm)  &  $\Lambda=0$ & $\Lambda=1/3$ & $\Lambda=2/3$ & $\Lambda=1$ \\ 
\hline
1   &  0.28  &  0.14 & 0.02  & 0.00 \\  
2   &  0.41  &  0.21 & 0.03  & 0.00 \\  
3   &  0.64  &  0.32 & 0.03  & 0.00 \\  
4   &  0.89  &  0.39 & 0.02  & 0.00 \\  
5   &  0.81  &  0.24 & 0.00  & 0.00 \\  
6   &  0.40  &  0.06 & 0.00  & 0.00 \\  
7   &  0.11  &  0.01 & 0.00  & 0.00 \\  
\hline

\end{tabular} 

(b)

  \begin{tabular}{ccccc} 
\hline 
\hline 
   $R$ (fm)  &  $\Lambda=0$ & $\Lambda=1/3$ & $\Lambda=2/3$ & $\Lambda=1$ \\ 
\hline
1   &  0.40  &  0.40 & 0.11  & 0.03 \\  
2   &  0.53  &  0.50 & 0.13  & 0.03 \\  
3   &  0.73  &  0.60 & 0.10  & 0.01 \\  
4   &  0.85  &  0.53 & 0.03  & 0.00 \\  
5   &  0.62  &  0.23 & 0.00  & 0.00 \\  
6   &  0.24  &  0.04 & 0.00  & 0.00 \\  
7   &  0.05  &  0.00 & 0.00  & 0.00 \\  
\hline

\end{tabular} 

\label{sqol-1-}
\end{table}

\section{Summary}\label{summary}

We have applied AQCM, which is a method 
to describe a transition from the $\alpha$-cluster wave function to the $jj$-coupling shell model
wave function, to $^{20}$Ne.
$^{20}$Ne has been known as a nucleus which has $^{16}$O+$\alpha$ structure, and
we investigated how the $\alpha$ cluster structure 
competes with independent particle motions of these four nucleons
by changing the strength of the spin-orbit interaction ($V_{ls}$).
We focused on the $E0$ transition matrix element, which was found to be sensitive to $V_{ls}$.

Based on AQCM,
$^{20}$Ne is characterized by only two parameters; 
$R$ representing the relative distance between $^{16}$O and $\alpha$
and $\Lambda$ describing the breaking of $\alpha$ cluster.
When the spin-orbit interaction is switched off ($V_{ls} = 0$ MeV),
the ground $0^+$ state has the squared overlap of 0.92
with the GCM basis state which has $R = 3$ fm and $\Lambda = 0$.
The second $0^+$ state is a higher nodal state and 
it has 0.74 with $R = 6$ fm and $\Lambda =0$.
The third $0^+$ state also has large $^{16}$O-$\alpha$ distance, and
the fourth $0^+$ state has overlaps with basis states with finite $\Lambda$ values.

When the spin-orbit interaction is switched on, 
we found that the decrease of the energy for the fourth $0^+$ state at $V_{ls} = 0$ is much steeper than other states.
Eventually the wave function of the fourth $0^+$ state at $V_{ls} = 0$ MeV 
strongly mixes in the ground and second $0^+$ states at $V_{ls} = 3000$ MeV.
On the other hand, the wave function of the second  $0^+$
state at $V_{ls} = 0$ MeV almost stays at this energy. 
Similar thing can be found for the third $0^+$ state at $V_{ls} = 0$ MeV.
These two states correspond to the third and fourth $0^+$ states at $V_{ls} = 3000$ MeV,
and $\alpha$ cluster structure becomes important again there.

The $E0$ transition matrix elements from the ground state 
to the second $0^+$ state
is calculated as 10.0 $e$ fm$^2$ at $V_{ls} = 0$ MeV,
which is slightly larger than
the experimental value (6.914 $e$ fm$^2$).
With increasing $V_{ls}$ value, the mixing of basis states with finite $\Lambda$ values
becomes important in both the ground and second $0^+$ states,
and the $E0$ transition matrix decreases.
The value agrees with the experimental one around $V_{ls} = 1770$ MeV.
This deduced strength is consistent with
our preceding work on the level structure of this nucleus.

At $V_{ls} = 1770$ MeV, which is the spin-orbit strength deduced from the present analysis on the
$E0$ transition matrix element,
the ground state has the squared overlap of 0.78 with the basis state 
which has $R = 3$ fm and $\Lambda =0$, and the character at $V_{ls} = 0$ MeV still remains.
However the largest squared overlap of 0.82 is with the basis state which has
$R = 2$ fm and $\Lambda = 1/3$. Therefore, the $\alpha$ breaking effect
due to the spin-orbit interaction is also important in the ground state.
The second $0^+$ states 
has squared overlap of 0.40 with the basis state which has $R = 5$ fm and $\Lambda = 0$;
however, 
it has also components of the basis states with finite $\Lambda$ values.
The third and fourth $0^+$ states are $\alpha$ cluster states and contain the components of
basis states with large $R$ values.

The presence of low-lying negative parity band starting with the first $1^-$
has been the key evidence for the $\alpha$ cluster structure.
We also investigated the $1^-$ states and found  that the energy of the first $1^-$ state 
is almost constant even if the spin-orbit interaction is switched on and
$\alpha$ breaking basis states are introduced.
The $\alpha$ cluster structure is really important in this state.

There have been discussions that the $^{12}$C+$\alpha$+$\alpha$ cluster states
appear in this energy region of the third $0^+$ state, and inclusion of this configuration
can be done by applying AQCM to the three $\alpha$ clusters in the $^{16}$O core.
Also, here we transformed an $\alpha$ cluster to four independent nucleons,
in which the spin-orbit interaction acts attractively.
However, in principle it is possible to introduce other shell model configurations,
for instance configurations where
the spin-orbit interaction acts repulsively, or one of the nucleon is excited from
$j$-upper orbit to $j$-lower orbit.
The analysis aiming at the unified view is going on.

\begin{acknowledgments}
Numerical calculation has been performed at Yukawa Institute for Theoretical Physics,
Kyoto University. 
\end{acknowledgments}


\begin{thebibliography}{100}
\bibitem{Horiuchi}
Hisashi Horiuchi and Kiyomi Ikeda, Prog. Theor. Phys. (1968) {\bf 40} 277.
%\bibitem{Zhou}
%Bo Zhou $et\ al.$, Phys. Rev. Lett. {\bf 110}, (2013) 262501.
\bibitem{Zhou}
Bo~Zhou, Zhongzhou~Ren, Chang~Xu, Y.~Funaki, T.~Yamada, A.~Tohsaki, H.~Horiuchi, P.~Schuck, and 
G.~R\"{o} pke, Phys. Rev. C {\bf 86}, 014301 (2012).
%\bibitem{Maris}
%P. Maris, J.P. Vary, and A.M. Shirokov, Phys. Rev. C {\bf 79}, 014308 (2009).
%\bibitem{Dreyfuss}
%Alison C. Dreyfussa, Kristina D. Launeyb, Tomas Dytrychb, Jerry P. Draayerb, Chairul Bahri,
%Phys. Lett. B {\bf 727}, 511 (2013).
%\bibitem{Yoshida}
%Tooru Yoshida, Noritaka Shimizu,
%Takashi Abe, and Takaharu Otsuka,
%J. Phys, Conf. Ser. {\bf 569}, 012063 (2014).
%\bibitem{Haxton}
%W.C. Haxton and Calvin Johnson, Phys. Rev. Lett. {\bf 65}, 1325 (1990).  
%\bibitem{Suzuki}
%Y. Suzuki, K. Arai, Y. Ogawa, and K. Varga, Phys. Rev. C {\bf 54}, 2073 (1996). 
%\bibitem{Mayer} 
%	M.G\H{o}ppert-Mayer, Phys. Rev. {\bf 75}, 1969 (1949).
%\bibitem{Jensen} 
%	O. Haxel, J.H.D. Jensen, and H.E. Suess, Phys. Rev. {\bf 75}, 1766 (1949).
%\bibitem{CSC}
%N. Itagaki, S. Aoyama, S. Okabe, and K. Ikeda,
%Phys. Rev. C {\bf 70}, 054307 (2004).
%\bibitem{Suhara2015}
%Tadahiro Suhara and Yoshiko Kanada-En'yo,
%Phys. Rev. C {\bf 91}, 024315 (2015).
\bibitem{Ne-Mg}
	N.~Itagaki, J.~Cseh, and M.~P{\l}oszajczak, Phys. Rev. C {\bf 83}, 014302 (2011).
\bibitem{Simple}
	N.~Itagaki, H.~Masui, M.~Ito, and S.~Aoyama, Phys. Rev. C {\bf 71} 064307 (2005). 
\bibitem{Masui}
	H.~Masui and N.~Itagaki, Phys. Rev. C {\bf 75} 054309 (2007). 
\bibitem{Yoshida}
	T.~Yoshida, N.~Itagaki, and T.~Otsuka, Phys. Rev. C {\bf 79} 034308 (2009).
\bibitem{Suhara}
     T.~Suhara, N.~Itagaki, J.~Cseh, and M.~P{\l}oszajczak,
Phys. Rev. C {\bf 87}, 054334 (2013).
\bibitem{General}
  N.~Itagaki, H.~Matsuno, and T.~Suhara, arXiv: 1507.02400.
\bibitem{Kawabata_PhysLettB646_6} T.~Kawabata \textit{et al}., Phys. Lett. B \textbf{646}, 6 (2007).
\bibitem{Sasamoto_ModPhysLettA21_2393} Y. Sasamoto \textit{et al}., Mod. Phys. Lett. A \textbf{21}, 2393 (2006).
\bibitem{Yoshida_PhysRevC79_034308} T.~Yoshida, N.~Itagaki, and T.~Otsuka, Phys. Rev. C \textbf{79}, 034308 (2009).
\bibitem{Yamada_PhysRevC92_034326}
Taiichi Yamada and Yasuro Funaki, Phys. Rev. C {\bf 92}, 034326 (2015).
\bibitem{Ichikawa_PhysRevC83_061301} T.~Ichikawa, N.~Itagaki, T.~Kawabata, Tz.~Kokalova and W.~von~Oertzen, Phys. Rev. C \textbf{83}, 061301(R) (2011).
\bibitem{Ichikawa_PhysRevC86_031303} T.~Ichikawa, N.~Itagaki, Y.~Kanada-En'yo, Tz.~Kokalova, and W.~von~Oertzen,
Phys. Rev. C \textbf{86}, 031303(R) (2012).
\bibitem{Matsuno}
H.~Matsuno and N.~Itagaki, arXiv: 1601.05892.
\bibitem{Brink}
	D.~M.~Brink, in {\it Proceedings of the International School of Physics "Enrico Fermi" Course XXXVI},
	edited by C. Bloch (Academic, New York, 1966), p. 247.
\bibitem{Elliot}
	J.~P.~Elliot, Proc. Roy. Soc. A {\bf 245} 128, 562 (1958)
\bibitem{Vol}
	A.~B.~Volkov, Nucl. Phys. {\bf 74}, 33 (1965).
\bibitem{G3RS} 
	R.~Tamagaki, Prog. Theor. Phys. {\bf 39}, 91 (1968).
\end{thebibliography}
\end{document}